\documentclass[onecolumn,amsmath,amssymb,pra]{revtex4}
\usepackage{setspace}
\usepackage{graphicx}
\usepackage{amsmath}
\usepackage{amssymb}
\usepackage{latexsym}

\begin{document}

\title{The decay and collisions of dark solitons in superfluid Fermi gases}
\author{R.G. Scott$^1$,  F. Dalfovo$^1$, L.P. Pitaevskii$^{1,2}$, S. Stringari$^1$, O. Fialko$^3$, R. Liao$^3$, J. Brand$^3$}
\affiliation{$^1$INO-CNR BEC Center and Dipartimento di Fisica, Universit{\`a} di Trento, Via Sommarive 14, I-38123 Povo, Italy.\\$^{2}$Kapitza institute for physical problems, ul. Kosygina 2, 119334 Moscow, Russia.\\$^3$New Zealand Institute for Advanced Study and Centre for Theoretical Chemistry and Physics, Massey University, Private Bag 102904 NSMC, Auckland 0745, New Zealand.} 



\begin{abstract}
We study soliton collisions and the decay of solitons into sound in superfluid Fermi gases across the Bose-Einstein condensate to Bardeen-Cooper-Schrieffer (BEC-BCS) crossover by performing numerical simulations of the time-dependent Bogoliubov-de Gennes equations. This decay process occurs when the solitons are accelerated to the bulk pair-breaking speed by an external potential. A similar decay process may occur when solitons are accelerated by an inelastic collision with another soliton. We find that soliton collisions become increasingly inelastic as we move from the BEC to BCS regimes, and the excess energy is converted into sound. We interpret this effect as being due to evolution of Andreev bound states localized within the soliton.
\end{abstract}

\maketitle

\section{Introduction}

Since the realization of Bose-Einstein condensation in dilute gases, experimental and theoretical research has shown that solitons play a key role in their dynamics (see Ref.~\cite{reviews} for recent reviews). Solitons have been generated in a wide variety of contexts such as, for example, phase-imprinting in elongated Bose-Einstein condensates (BECs)~\cite{bongs,Burger}, transport in optical lattices~\cite{RScottOL,RScottOL2}, transport past single defects or through disordered potentials~\cite{leboeuf,hulet,engels}, BEC collisions~\cite{collision3,burnett,Steinhauer} and interferometry~\cite{kettnew,RScottInter}. Consequently, solitons might be expected to play an equally important role in the dynamics of degenerate Fermi gases. Recent work has predicted the existence of black~\cite{Antezza} and grey solitons~\cite{metrento,brand,carrpaper} across the Bose-Einstein condensate to Bardeen-Cooper-Schrieffer (BEC-BCS) crossover, and that they may perform stable oscillations in a trap~\cite{metrento}. However, if solitons in Fermi gases are to be as ubiquitous as solitons in Bose gases, they must be robust in the presence of defects, rapidly-varying potentials, superfluid flow, sound and other solitons. Can they be easily produced and are they long-lived? If not, what excitations will be observed?

In this paper, we address some of these questions by solving the time-dependent Bogoliubov-de Gennes (TDBdG) equations across the BEC-BCS crossover. We confirm a previous hypothesis that soliton solutions do not exist with a speed $v$ above the bulk pair-breaking speed $v_{pb}$~\cite{brand}. This means that on the BCS side of unitarity, where $v_{pb}$ is smaller than the sound speed $c$, the maximum $v$ is $v_{pb}$, in accordance with the Landau criterion. In contrast, in the BEC regime the maximum $v$ is $c$, as predicted by the Gross-Pitaevskii equation~\cite{levsandro,pethick}. If a soliton in the BCS regime is accelerated to $v_{pb}$ by an external potential, it abruptly disappears and its energy is converted into sound. We stress that this decay process is distinct from the snake instability~\cite{feder,anderson,brandsnake,dutton}, by which a soliton can decay into quantized vortices in a three-dimensional superfluid. In our calculations, in fact, we assume that the dynamics are restricted to longitudinal motion only, the gas remaining uniform in the transverse directions. This is equivalent to assuming that, in a real experiment, the superfluid is sufficiently tightly-confined in the transverse directions to suppress the snake instability. Moreover, the predicted decay process occurs at zero temperature and is distinct from other decay mechanisms considered previously~\cite{janne,cockburn,gangardt}.

We analyse this decay process by looking at the low-lying eigenstates of the Bogoliubov quasi-particle spectrum. The lowest state is found to be localized in the vicinity of the soliton, with an energy below the gap for the continuous spectrum of the extended Bogoliubov states. The origin of this localized state (Andreev state~\cite{andreev,andreev2,james}) is the fact that the energy cost for creating a fermionic excitation near a minimum of the order parameter is reduced with respect to the bulk value. The Andreev states associated with dark solitons at rest have already been discussed in Ref.~\cite{Antezza}. Here we find that Andreev states also exist for solitons with finite $v$. However, as $v$ approaches $v_{pb}$, the excitation gap in the continuous spectrum approaches zero and the Andreev bound state is lost. We also study the soliton energy $E_s$ as a function of $v$ (the energy dispersion), and find that the soliton still has a finite $E_s$ and phase jump across it at the maximum $v$. The energy dispersion shows interesting structure as $v$ approaches $v_{pb}$, in particular a local minimum. We explain that, in a real experiment, this local minimum would cause the soliton to decay at the value of $v$ in the minimum rather than at the maximum $v$ where the energy dispersion truncates. Moreover, the presence of a harmonic trap causes a further reduction in the value of $v$ at which the soliton decays. This finite-size effect is reduced as we weaken the trap.

The existence of a maximum soliton speed, or minimum soliton energy, has important consequences for the physics of soliton collisions across the crossover. In the BEC limit, due to the integrability of the one-dimensional Gross-Pitaevskii equation for a homogeneous gas, soliton collisions are elastic and hence solitons always emerge from a collision with the same velocity and energy at which they collided~\cite{collision1,stellmer}. However, we find that soliton collisions become increasingly inelastic as we reduce $1/k_f a$. As for the decay of a single soliton, fermionic quasi-particles are found to play a key role also in these inelastic collisions. This is consistent with the fact that the same collisions between solitons are found to be almost completely elastic~\cite{wenwen} when described by means of a nonlinear Schr{\"o}dinger equation, which is often used to investigate the dynamics of superfluid fermions in the BEC-BCS crossover, but including only bosonic degrees of freedom. Since, counter-intuitively, fast solitons have less energy than slow solitons~\cite{levsandro,pethick,brand}, the inelastic collisions cause the solitons to \emph{accelerate}. The energy lost by the solitons is dissipated as sound. We interpret this effect as being due to the evolution of the Andreev bound states as the soliton changes shape during the collision; this evolution eventually causes an energy transfer from the soliton to the continuum of bulk excitations. If the solitons lose so much energy that their energy after the collision would be less than the minimum soliton energy, they are destroyed by the collision.

\section{Methodology}

We consider a three-dimensional superfluid Fermi gas with equal populations of the two spin components. We model its dynamics across the BEC-BCS crossover by solving the TDBdG equations~\cite{metrento,Challis}
\begin{equation}
\left[\begin{array}{ll}
\hat{H} & \Delta \mbox(\textbf{r},t) \\
\Delta^* \mbox(\textbf{r},t) & -\hat{H}
\end{array}\right]
\left[\begin{array}{l}
u_\eta \mbox(\textbf{r},t) \\
v_\eta \mbox(\textbf{r},t)
\end{array}\right] = 
i \hbar \frac{\partial}{\partial t}
\left[\begin{array}{l}
u_\eta \mbox(\textbf{r},t) \\
v_\eta \mbox(\textbf{r},t)
\end{array}\right] . 
\label{eq:tdbdg}
\end{equation}
where $\hat{H} = -\hbar^2 \nabla^2 / 2m + U - \mu\left(0\right)$, in which $m$ is the atomic mass, $U$ is the external potential and $\mu$ is the chemical potential. The order parameter is calculated as $\Delta\left(\textbf{r},t\right) = -g \sum_{\eta}u_{\eta}v_{\eta}^{*}$, in which $g$ is given by $1/k_f a = 8\pi E_f/ ( g k_f^3 ) + \sqrt{4E_c / \left(\pi^2 E_f\right)}$~\cite{SandroReview}. Here $a$ is the 3D s-wave scattering length characterizing the interaction between atoms of different spins, while $E_f = \hbar^2 k_f^2 / 2m$ and $k_f = \left(3\pi^2n\right)^{1/3}$ are the Fermi energy and momentum of an ideal Fermi gas of density $n$, respectively. The cut-off energy $E_c$ is introduced in order to remove the ultraviolet divergences in the BdG equations with contact potentials. The density of the gas is $n \mbox(\textbf{r},t) = 2 \sum_{\eta} \left|v_{\eta} \mbox(\textbf{r},t)\right|^2$. We impose that the potential $U$ has no $y$ or $z$ dependence, and hence we may write the functions $u_\eta \mbox(\textbf{r},t)$ and $v_\eta \mbox(\textbf{r},t)$ as $u_\eta (x,t) e^{i (k_y y + k_z z)}$ and $v_\eta (x,t) e^{i (k_y y + k_z z)}$ respectively, in which $k_y$ and $k_z$ are quantized according to $k_y = 2\pi \alpha_y / L_\bot$ and $k_z = 2\pi \alpha_z / L_\bot$, where $\alpha_y$ and $\alpha_z$ are integers and $L_\bot$ is the width of the box in the $y$- and $z$-directions. As initial states at $t=0$, we find stationary solutions of Eq.~(\ref{eq:tdbdg})~\cite{metrento,Antezza}.

We also search for solutions of Eq. (\ref{eq:tdbdg}) satisfying $\Delta\left(x,t\right) = \Delta\left(x - vt\right)$, given that $U=0$, by solving the equation
\begin{equation}
\left[\begin{array}{ll}
\hat{H}_\xi+i\hbar v \frac{d}{d\xi} & \Delta(\xi) \\
\Delta^*(\xi) & -\hat{H}_\xi+i\hbar v \frac{d}{d\xi}
\end{array}\right]
\left[\begin{array}{l}
u_\eta(\xi) \\
v_\eta(\xi)
\end{array}\right] = 
\epsilon_\eta
\left[\begin{array}{l}
u_\eta(\xi) \\
v_\eta(\xi)
\end{array}\right] , 
\label{eq:tibdg}
\end{equation}
where $\xi = x - vt$, $\hat{H}_\xi = -\hbar^2 / 2m \left[ \partial^2 / \partial\xi^2 - k_y^2 - k_z^2 \right] - \mu$ and $\epsilon_\eta$ is the energy of level $\eta$. We use a generalized secant (Broyden's) method to find self-consistent solutions~\cite{brand}. Hence we find travelling soliton solutions in a homogeneous gas which are stationary in the frame of the soliton, thus generalising to the BEC-BCS crossover the solutions for solitons propagating in a homogeneous Bose gas~\cite{tsuzuki,levsandro,pethick}. This technique enables us to determine the soliton properties more accurately in the homogeneous gas and identify effects due to the trapping potential by comparison with the time-dependent simulations.


\section{Soliton decay}

Some of us have shown in a previous publication~\cite{brand} that the energy dispersion of the soliton in the BCS regime truncates below the speed of sound $c$. This result suggested that, more generally, soliton solutions do not exist for speeds greater than $v_{pb}$ or $c$, whichever is the smaller. To support this hypothesis, and to show how the effect would manifest itself in a real experiment, we now present time-dependent simulations of the Bogoliubov-de Gennes equations~\cite{Challis,metrento} in which we accelerate the soliton above the critical velocity.  

\begin{figure}[tbp]
\includegraphics[width=0.6\columnwidth]{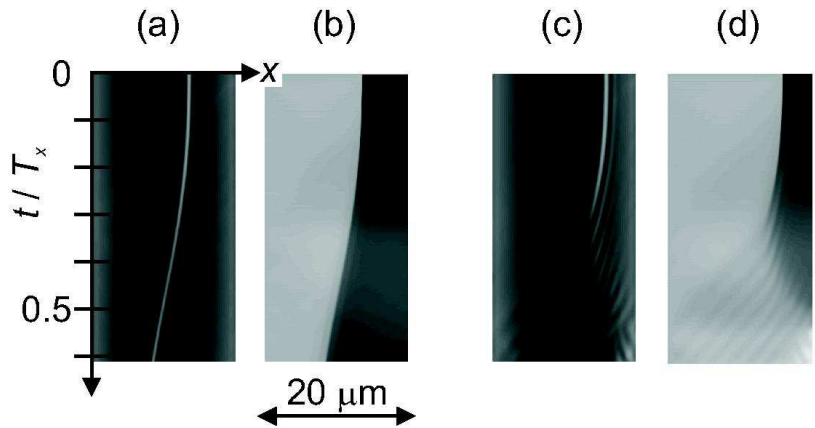} 
\caption{(a) Grey-scale plot of density profile $n(x,t)$ (black=high) and (b) phase of the order parameter $\Delta(x,t)$ for a stable soliton oscillation at $1/ k_f a = -0.5$ in a $^{40}K$ superfluid, with $\omega_x = 2 \pi \times 50$ rad s$^{-1}$, $L_\bot = 3.3$ $\mu$m and a peak density $n_p = 1.8 \times 10^{18}$ m$^{-3}$. The soliton begins at rest at a distance $X_0 = 3.3$ $\mu$m from the trap centre. (c) \& (d): Corresponding plots for a larger $X_0$ of $5.4$ $\mu$m producing soliton decay.}
\label{f0}
\end{figure}

Firstly, for comparison, in Fig.~\ref{f0} we present a simulation of a stable soliton oscillation in a trap at $1/ k_f a = -0.5$. As in a previous publication~\cite{metrento}, we consider a soliton oscillating in a $^{40}K$ superfluid, contained in the harmonic trapping potential $U(x) = m \omega_x^2 x^2 /2$, with $\omega_x = 2 \pi \times 50$ rad s$^{-1}$, $L_\bot = 3.3$ $\mu$m and a peak density $n_p = 1.8 \times 10^{18}$ m$^{-3}$. Figures~\ref{f0}(a) and (b) show the density profile and phase of the order parameter. Note that only the region of cloud near the center of the trap is shown, the low density tails of the cloud are outside of the field-of-view. The soliton begins at rest at a distance $X_0 = 3.3$ $\mu$m from the trap centre, and is consequently accelerated by the harmonic potential. As this happens, the density profile of the soliton becomes shallower and the phase jump across the soliton reduces from $\pi$. However, as the soliton passes the center of the trap and reaches its maximum speed, it is still localized in the density profile and the phase jump is well-defined. The soliton then begins to decelerate as it climbs the trap potential, and the phase jump increases towards $\pi$. If the simulation were allowed to continue the soliton would perform a complete oscillation in the harmonic potential.

Figures~\ref{f0}(c) and (d) show the corresponding density profile and phase for the larger $X_0$ of $5.4$ $\mu$m. The soliton begins to accelerate, and initially the soliton remains localized in the density profile with a clear phase jump. However, when the soliton reaches a critical speed it rapidly spreads out in the density profile and the phase jump disappears. At the end of the simulation we see only low-amplitude modulations of the density profile. By measuring their speed we identify these modulations as sound.

\begin{figure}[tbp]
\includegraphics[width=0.6\columnwidth]{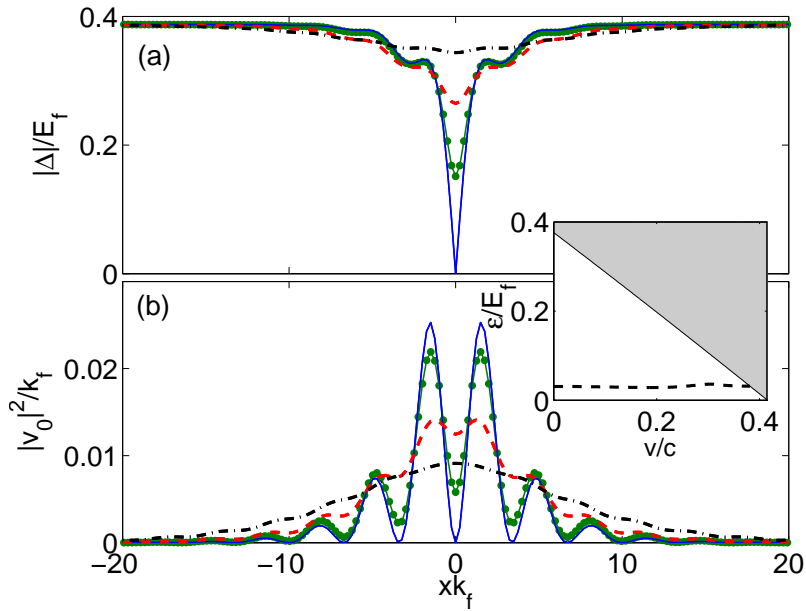} 
\caption{(a) Profile of the modulus of the order parameter $\left|\Delta(x)\right|$ for $1/k_f a = -0.5$ and $v=0$ (solid blue curve), $0.2c$ (solid green curve with cicles), $0.3c$ (dashed red curve) and $0.38c$ (dot-dashed black curve), obtained from the time-independent calculations of a homogeneous superfluid. (b) Corresponding plots of the lowest Andreev bound state (specifically $\left|v_0(x)\right|^2$). Inset: the spectrum of energy levels in a homogeneous gas containing a soliton. The solid line shows the excitation gap in the continuous spectrum of the extended Bogoliubov states in a homogeneous Fermi gas as a function of velocity for $1/k_f a = -0.5$. The shaded area above represents the continuum spectrum for the homogeneous infinite system. The dashed line shows the energy of the lowest energy level, which is the Andreev bound state shown in (b).}
\label{spectrum}
\end{figure}

To gain a deeper physical understanding of this effect, we now study the soliton in the homogeneous gas using the BdG equations (\ref{eq:tibdg}). The magnitude of the order parameter $\left|\Delta(x)\right|$ is plotted in Fig.~\ref{spectrum}(a) for $1/k_f a = -0.5$ and $v=0$ (solid blue curve), $0.2c$ (solid green curve with circles), $0.3c$ (dashed red curve) and $0.38c$ (dot-dashed black curve). As $v$ approaches the pair-breaking velocity $v_{pb} = 0.41 c$, the soliton becomes very shallow and broad. This is reflected in the profile of the lowest Andreev state, shown in Fig.~\ref{spectrum}(b). At $v=0$, $\left|v_0(x)\right|^2$ is tightly localised at the soliton position~\cite{Antezza} (blue solid curve). Also note that there are many nodes in the function. As $v$ increases to $0.2c$ (green solid curve with circles) and $0.3c$ (red dashed curve), the function widens and the nodes becomes gentle oscillations. When $v$ reaches $0.38c$, the function is broad and smooth with almost no discernable oscillations (black dot-dashed curve). 

As $v$ approaches the pair-breaking velocity $v_{pb} = 0.41 c$ for $1/k_f a = -0.5$, the excitation gap in the continuous spectrum of the extended Bogoliubov states, shown by the black solid line in the Fig.~\ref{spectrum} inset, approaches zero. Below the continuous spectrum, there is the single Andreev bound state, which is plotted for various $v$ in Fig.~\ref{spectrum}(b). As the excitation gap closes, the lower boundary of the continuous spectrum approaches the energy of the Andreev bound state, shown by the dashed line in the Fig.~\ref{spectrum} inset. Note that the energy of the Andreev bound state is almost independent of velocity. Eventually, the localized Andreev bound state is lost and becomes an extended state. 


The loss of the Andreev state can be understood by a simple analytic argument. We treat Eq. (\ref{eq:tibdg}) semiclassically by considering the situation where the order parameter $\Delta$ varies slowly over scales of the order of $k_f^{-1}$ \cite{andreev}. This is true for the soliton near its maximum velocity, as shown in Fig.~\ref{spectrum}(a). This assumption allows us to decouple Eq. (\ref{eq:tibdg}) into two separate second-order differential equations for the amplitudes $u_0=f(z) \exp (-ik_fz)$ and $v_0=g(z) \exp (-ik_fz)$, by neglecting terms proportional to $\partial^2_z f$ compared to terms in $k_f\partial_z f$, and the same for $g$. Each of the resulting equations has two independent solutions of the form $\exp[\pm i k z]$, where $k\sim [v_f^2(\epsilon_0-\hbar k_f v)^2-\Delta_0^2(v_f^2-v^2)]^{1/2}$ with $v_f = \hbar k_f /m$ being the Fermi velocity, $\epsilon_0$ being the energy of the Andreev bound state [the dashed line in the Fig.~\ref{spectrum} inset], and $\Delta_0$ being the bulk order parameter. The expression under the square root is zero at the pair-breaking velocity $v\approx v_{pb} \approx v_f|\Delta_0|/2\mu$ in the BCS regime. For smaller velocities $k$ is imaginary and hence the Andreev bound state is localized. The size of the bound state is thus $l_b\sim 1/|k|$, which diverges at the pair-breaking velocity. Our calculations suggest that the soliton solution ceases to exist at a velocity slightly below $v_{pb}$ where the excitation gap is still finite.

\begin{figure}[tbp]
\includegraphics[width=0.6\columnwidth]{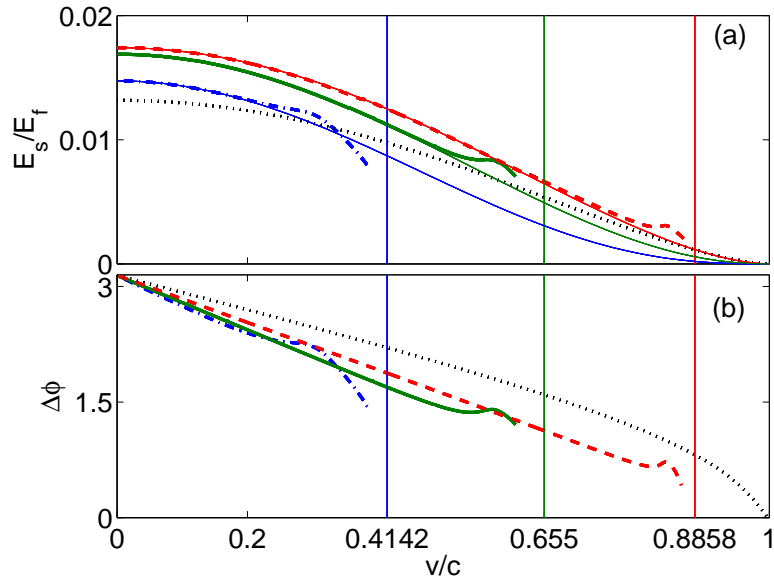} 
\caption{a) Energy of soliton $E_s$ in units of the Fermi energy $E_f$ as a function of soliton velocity $v$ for $1/ k_f a = -0.5$ (blue dot-dashed curve), $-0.2$ (green solid curve), $0$ (red dashed curve) and $1$ (black dotted curve), obtained from the time-independent calculations of a homogeneous superfluid. The curves of numerical results for $1/ k_f a = -0.5$, $-0.2$ and $0$ are fitted to thin solid curves of the form $E_s \propto \left(1-v^2/c^2\right)^{\alpha}$. The solid horizontal lines indicate $v_{pb}$ for $1/ k_f a = -0.5$, $-0.2$ and $0$ from left to right. (b) Corresponding plot for the phase jump across the soliton $\Delta\phi$.}
\label{franco}
\end{figure}

In Fig.~\ref{franco}(a) we plot the energy of the soliton $E_s$ as a function of $v$ (the energy dispersion) for various values of $1/k_f a$. In the BEC limit, the energy is given by the well-known Gross-Pitaevskii result~\cite{tsuzuki,levsandro,pethick} 
\begin{equation}
E_s \propto \left(1-v^2/c^2\right)^{\alpha} , 
\label{eq:fit}
\end{equation}
where $\alpha=3/2$. For $1/k_f a=1$ (black dotted curve), our numerical data closely follows this result. At unitarity (red dashed curve), the energy initially decreases smoothly with increasing $v$, as for $1/k_f a=1$, but then reaches a local minimum, subsequently briefly increases, before decreasing again rapidly. Finally the energy dispersion truncates at a small but finite value of $E_s$~\cite{note1}. In a previous publication, some of us predicted that at unitarity $\alpha=2$ in Eq. (\ref{eq:fit})~\cite{brand}. This is approximately true for small $v$. The presence of a local minimum in the energy dispersion is a crucial point, as we now explain.

In general, the soliton energy $E_s$ is a function of $\mu$ and $v^2$~\cite{metrento}. In Fig~\ref{f0}(c), the soliton begins at a distance $X_0$ from the trap center with $v=0$. The soliton may move closer to the center of the trap, to larger $\mu$, maintaining a constant $E_s$, by increasing $v^2$. However, once the soliton reaches the local minimum in the energy dispersion, it can no longer move to larger $\mu$ by changing $v^2$. Hence, in a real experiment, the soliton must decay when $v$ reaches the position of the local minimum, \textit{not} when it reaches the point where the curve truncates.

Following this line of reasoning, we may reach another important conclusion. If the energy of the soliton were to decrease monotonically to zero with increasing $|v|$, as in the Gross-Pitaevskii equation, decay of the kind shown in Fig.~\ref{f0}(c) would be impossible. This is because the soliton could always maintain a constant $E_s$, however much $\mu$ increased, by increasing $v$ closer to the maximum $v$ where $E_s$ goes to zero. Hence, the existence of the decay process shown in Fig.~\ref{f0}(c) implies a truncation in the energy dispersion, or at least a non-monotonic energy dispersion.



When we reduce $1/k_f a$ from $0$ to $-0.2$ [green solid curve in Fig.~\ref{franco}(a)], we again find a local minimum in the energy dispersion, but the curve truncates at a much larger value of $E_s$. We now obtain $\alpha\simeq2.2$ in Eq. (\ref{eq:fit}). The minimum value of $E_s$ is even larger for $1/k_f a=-0.5$ (blue dot-dashed curve), and $\alpha$ has increased to $2.8$, but now the energy dispersion does not contain a minimum. Instead, we see a broad bump in the curve as the soliton approaches the maximum velocity. 


The non-monotonic slope of the energy dispersion in the BCS regime can be tentatively explained by considering the Friedel oscillations. These arise due to scattering of fermions at the Fermi energy by the soliton, and appear as ripples in the density or order parameter profile either side of the central minimum of the soliton [as shown in Fig.~\ref{f0}(a), for example]. For Friedel oscillations to appear, the atomic wave length at the Fermi energy, $\lambda \approx 2\pi/k_f$, should be larger than the size of the soliton. When the velocity of the soliton increases, its size grows and can exceed $\lambda$, causing the loss of the Friedel oscillations. For example, we find that the size of the bound state $l_b$ and $\lambda$ are comparable at $v = 0.3 c$ for $1/k_f a = -0.5$. For $v=0$, $0.2c$ and $0.3c$, the Friedel oscillations are almost identical in the profile of the order parameter [blue solid curve, green solid curve with circles and red dashed curve in Fig.~\ref{spectrum}(a), respectively], although the depth of the soliton varies dramatically. In contrast, if $v$ increases to $0.38c$, the Friedel oscillations disappear and the order parameter acquires a smooth broad profile [black dot-dashed curve in Fig.~\ref{spectrum}(a)]. Due to the disappearance of the Friedel oscillations, we would expect different behaviour in the energy dispersion, as observed in Fig.~\ref{franco}(a). Notice that the existance of a small range of $v$ where $E_s$ increases with $v$ implies that, if a soliton were created by phase imprinting in this range of $v$, the snake instability would be absent~\cite{levsnake}.

Figure \ref{franco}(b) shows the equivalent plot to Fig.~\ref{franco}(a) for the phase jump across the soliton $\Delta\phi$. In the BEC limit, the Gross-Pitaevskii equation predicts that $\Delta\phi = -2 \arccos \left(v/c\right)$. For $1/k_f a=1$ (black dotted curve), our numerical data again agrees well with the Gross-Pitaevskii prediction. At unitarity (red dashed curve), $\Delta\phi$ varies linearly with $v$ for small $v$, as some of us predicted analytically in a previous publication~\cite{brand}. However, as $v$ approaches the maximum velocity we see a departure from this linear behaviour: in a similar manner to the energy dispersion, the $\Delta\phi(v)$ curve forms a local minimum, then briefly rises before decreasing rapidly and terminating. This plot makes the important point that the soliton still has a finite phase jump when the energy dispersion terminates. This is expected because, as shown in Fig.~\ref{franco}(a), the energy dispersion truncates at finite energy. In the case of a soliton moving with a nonzero velocity $v$, a finite phase step implies finite energy by virtue of the kinetic energy associated with a superfluid current. When we decrease $1/k_f a$ from $0$ to $-0.2$ we find a similar variation of $\Delta\phi$ with $v$ (green solid curve), except that the curve now truncates at a larger $\Delta\phi$. As for the energy dispersion, at $1/k_f a=-0.5$ (blue dot-dashed curve) we find a broad bump instead of the local minimum.

\begin{figure}[tbp]
\includegraphics[width=0.5\columnwidth]{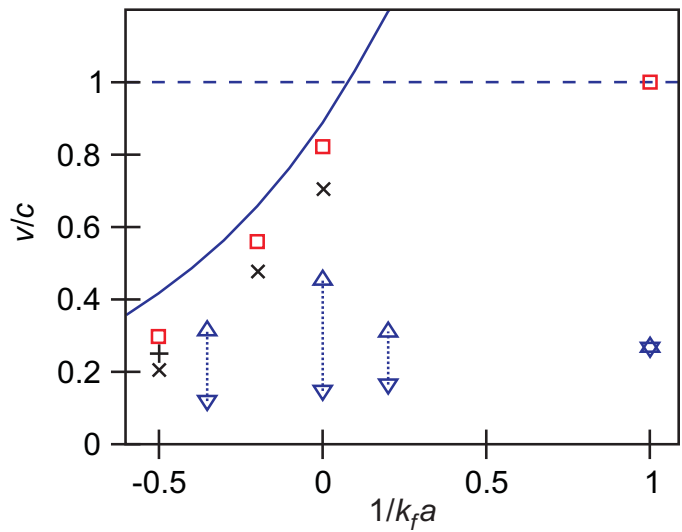} 
\caption{Dashed curve: sound speed $c$ in a homogeneous superfluid as a function of $1/ k_f a$. Solid curve: bulk pair-breaking velocity $v_{pb}$. Red squares: maximum soliton speed $v_{pb}$ obtained from the time-independent calculations of a homogeneous superfluid. Green crosses (plus sign): maximum soliton speed obtained from time-dependent calculations of a superfluid in a trap with frequency $\omega_x = 2\pi \times 50$ ($2\pi \times 25$) rad s$^{-1}$. Blue downward (upward) pointing triangles: soliton speeds before (after) the collisions presented in the middle row of Fig.~\ref{f1}.}
\label{vc}
\end{figure}

Using the above results we plot the maximum soliton velocity $v_m$ in the homogeneous gas as a function of $1/k_f a$ as red squares in Fig.~\ref{vc}. As explained earlier, in a real or numerical experiment, we predict that the measured $v_m$ is given by the position of the local minimum in the energy dispersion, \textit{not} by the the point where the curve truncates, and hence we define $v_m$ to be the position of that minimum. For $1/k_f a=-0.5$ there is no local minimum in the energy dispersion, but we take $v_m$ to be the velocity at the ``broad bump'' for consistency. We compare this data to the bulk sound speed $c$ (dashed curve) and bulk pair-breaking velocity $v_{pb}$ (solid curve) for a homogeneous gas, given by $m v_{pb}^2 = \sqrt{\Delta^2 + \mu^2} - \mu$. For $1/k_f a=1$, the maximum soliton velocity is given by $c$, as expected from the Gross-Pitaevskii equation~\cite{levsandro}. The pair-breaking speed is much larger and hence not relevant because the pairs are strongly bound as molecules. However, for $1/k_f a\leq0$, the pair-breaking speed determines the Landau critical velocity, and hence limits $v_m$. Thus, at unitarity we find a $v_m$ which is slightly smaller than $v_{pb}$. As we decrease $1/k_f a$ we find that $v_m$ closely follows the $v_{pb}$ curve. In fact, the data points lie slightly below $v_{pb}$, and the deviation becomes larger as $1/k_f a$ decreases. 

On the same graph we also include a number of points (black crosses) showing the critical velocity predicted by the time-dependent simulations of a trapped superfluid. The points lie slightly below the predictions of the time-independent calculations, the relative deviation being less at unitarity than at smaller values of $1/k_f a$. This small deviation is a finite-size effect, which probably occurs because the pair size for small $1/k_f a$ becomes comparable to the size of the cloud. Hence it would be exceedingly difficult to probe soliton dynamics near the bulk pair-breaking speed in a real experiment. We include one point (green plus sign) in Fig.~\ref{vc} for $\omega_x = 2 \pi \times 25$ rad s$^{-1}$ and $1/k_f a = -0.5$ to illustrate that the critical velocity in the time-dependent simulations approaches the prediction of the time-independent calculations as the trap frequency is reduced.

\section{Soliton collisions}

\begin{figure}[tbp]
\includegraphics[width=0.6\columnwidth]{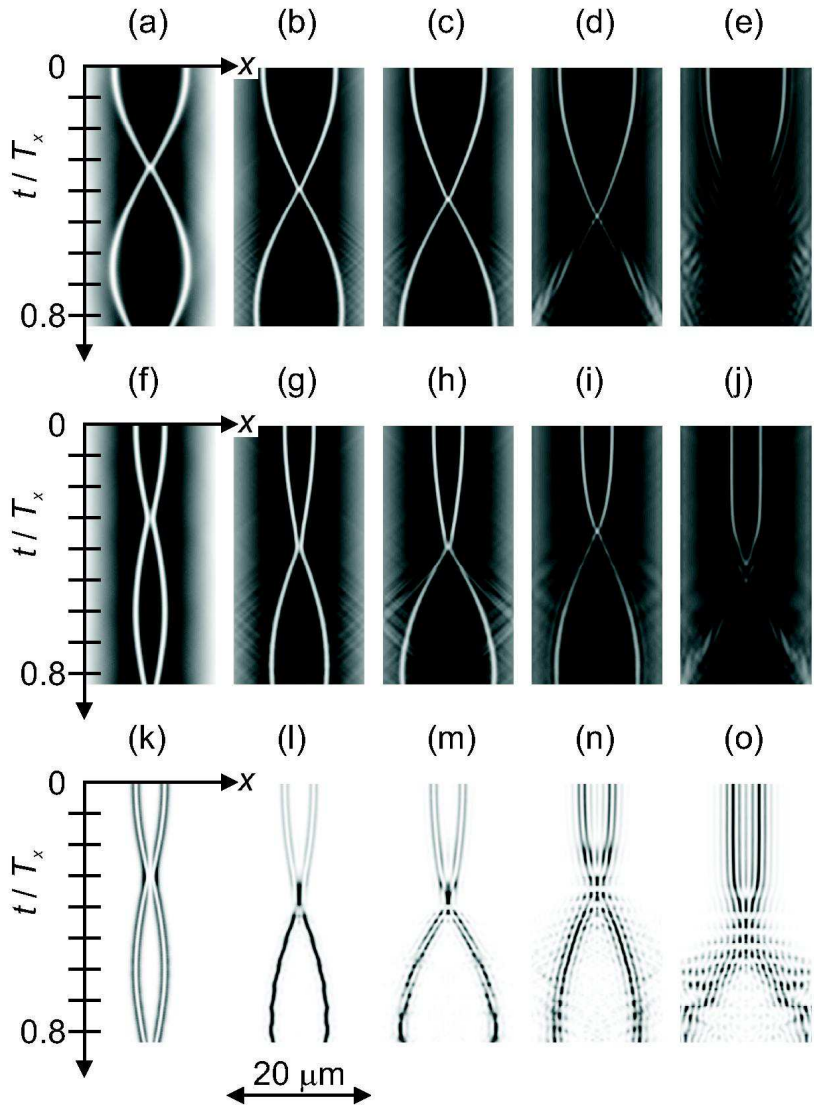} 
\caption{Collisions of solitons across the BEC-BCS crossover. The top (middle) row shows grey-scale plots of the density profile $n(x,t)$ (black=high) in a $^{40}K$ superfluid, in which two solitons begin at rest with a displacement $X_0 = \pm 5.1$ ($2.1$) $\mu$m from the trap centre, $\omega_x = 2 \pi \times 50$ rad s$^{-1}$, $L_\bot = 3.3$ $\mu$m and the peak density $n_p = 1.8 \times 10^{18}$ m$^{-3}$. The bottom row shows the evolution of the lowest Andreev state (specifically $\left|v_0(x,t)\right|^2$) in grey-scale (black=high) during the collisions shown in the middle row. $1/ k_f a = 1.0$, $0.2$, $0.0$, $-0.35$ and $-0.5$ in each column from left to right.}
\label{f1}
\end{figure}

We place two black solitons in a trapped Fermi superfluid, with a displacement of $\pm X_0$ from the trap centre, and then evolve in time. The solitons are accelerated by the harmonic potential and collide at the trap centre. We may increase the speed of the collision by increasing $X_0$. Figure \ref{f1} shows the evolution of the density profile with time for different $X_0$ and $1/k_f a$, with the other parameters as in Fig.~\ref{f0}. In the upper (middle) row $X_0 = 5.1$ ($2.1$) $\mu$m, and $1/k_f a$ decreases from left to right (see caption). 

Let us begin with the BEC regime $1/k_f a = 1$, shown in the left-hand column. For $X_0 = 5.1$ $\mu$m [Fig.~\ref{f1}(a)], the two solitons pass through each other without changing their form, and subsequently slow down and come to rest at $x \approx \pm X_0$, indicating that no energy has been lost by the solitons and hence the collision is elastic. For $X_0 = 2.1$ $\mu$m [Fig.~\ref{f1}(f)], the solitons behave differently. In this case, the solitons slow down as they approach one another, come to a halt, and then move away without crossing. This is known as a black collision, whereas a crossing of two solitons, as shown in Fig.~\ref{f1}(a), is known as a grey collision. We again note that the two solitons return to their original positions after the collision, indicating that the collision is elastic. This behaviour of grey collisions for high impact speeds, and black collisions for low impact speeds, has already been observed in the GP equation~\cite{collision1}. 

When we reduce $1/k_f a$ to $0.2$, the behaviour is similar. We again observe a grey collision for $X_0 = 5.1$ $\mu$m [Fig.~\ref{f1}(b)], and a black collision for $X_0 = 2.1$ $\mu$m [Fig.~\ref{f1}(g)]. However, on close inspection we observe that, after the collision, the solitons come to rest at a value of $|x| > X_0$, indicating that the solitons have a greater speed after the collision than before. This is particularly noticeable for the slower collision [Fig.~\ref{f1}(g)]. Since, counter-intuitively, fast solitons have less energy than slow solitons (as shown in Fig~\ref{franco}), we conclude that the solitons have lost energy due to an inelastic collision. We also observe small ripples in the density which emanate from the collision, showing that the energy lost by the solitons is converted into sound. 


We interpret this effect as being due to the evolution of fermionic quasiparticles localised in the solitons. In the bottom row of Fig.~\ref{f1}, we plot the evolution of the lowest Andreev state (specifically $\left|v_0(x,t)\right|^2$) during the slow soliton collisions shown in the middle row of Fig.~\ref{f1}. As the solitons approach, the Andreev states see a double-well potential, and subsequently a deep single-well (if the collision is grey), causing them to oscillate and breathe in a non-adiabatic way. In the BEC regime this effect is negligible, because the Andreev states make a small contribution to the overall density ($\int u_0 \: dx \: dy \: dz \gg \int v_0 \: dx \: dy \: dz$, for example), and because the solitons repel each other at slow speeds. Consequently the slow collision causes no disruption to the lowest Andreev state for $1/k_f a = 1.0$ [Fig.~\ref{f1}(k)]. However, for $1/k_f a = 0.2$ we observe oscillations in the amplitude of the lowest Andreev state [Fig.~\ref{f1}(l)]. At one instant, shortly after the collision, $\left|v_0(x,t)\right|^2$ reduces almost to zero. This means that particles are being transferred between different eigenstates of the Bogoliubov spectrum, with a coupling between the localized Andreev state and the states in the continuum. These transitions are associated with density and phase oscillations which eventually cause an emission of sound and loss of energy from the soliton. This effect is enhanced for low impact-collisions, because the fermionic quasiparticles have more time to move in response to the change in potential. The key role played by fermionic quasiparticles in this process is also confirmed by the fact that the same collisions are found to be elastic when a purely bosonic density functional theory (i.e., a nonlinear Schr{\"o}dinger equation for the pairing field) is used in place of the TDBdG equations~\cite{wenwen}.

As we further reduce $1/k_f a$, the collisions become increasingly inelastic. At unitarity [Figs.~\ref{f1}(c) and (h)], the collisions create a great deal of sound and the solitons subsequently come to rest at much larger $|x|$ than $X_0$, particularly for $X_0 = 2.1$ $\mu$m [Fig.~\ref{f1}(h)]. The inelastic collision is associated with very pronounced oscillations in $\left|v_0(x,t)\right|^2$ [Fig.~\ref{f1}(m)]. We also observe that the collision for $X_0 = 2.1$ $\mu$m is no longer black, due to the increasing mass of the solitons~\cite{metrento}. When $1/k_f a$ reaches $-0.35$, the solitons are destroyed by the collision for $X_0 = 5.1$ $\mu$m [Fig.~\ref{f1}(d)], but survive if $X_0 = 2.1$ $\mu$m [Fig.~\ref{f1}(i)]. Solitons can be destroyed by the inelastic collisions if their residual energy after the collision is less than the minimum soliton energy~\cite{brand0}. For $1/k_f a = -0.5$ and $X_0 = 5.1$ $\mu$m, the minimum energy (or maximum speed) is reached before the solitons even collide [Fig.~\ref{f1}(e)]. For $X_0 = 2.1$ $\mu$m, the solitons collide but then are immediately destroyed [Fig.~\ref{f1}(j)]. Consequently the lowest Andreev state is no longer localised after the collision [Fig.~\ref{f1}(o)].

We quantify how inelastically the solitons collide in Fig.~\ref{vc}. Here we plot the soliton speed immediately before and after the soliton collisions presented in the middle row of Fig.~\ref{f1} as pairs of downward and upward-pointing triangles respectively. For $1/k_f a=1$, the collision is elastic, and hence the downward and upward-pointing triangles lie on top of each other. For $1/k_f a=0.2$, the triangles are a small distance apart, indicating a slightly inelastic collision. This distance increases as $1/k_f a$ is reduced to $0$, showing that the collision is becoming increasingly inelastic. However, the speed of the soliton after the collision is still far below the maximum speed observed for a single soliton, shown by the black crosses. However, when $1/k_f a$ is reduced to $-0.35$ the final soliton speed is very close to the critical value. This explains why collisions for larger $X_0$ or smaller $1/k_f a$ destroy the solitons.

\section{Conclusions}

We have shown that in a two-component superfluid Fermi gas soliton solutions cease to exist as the soliton approaches the bulk pair-breaking velocity $v_{pb}$ on the BCS side of the resonance. At this point, the variation of soliton energy with velocity (the energy dispersion) truncates at a finite value of energy and phase jump across the soliton. Close to $v_{pb}$, the energy dispersion shows interesting structure, and sometimes a local minimum, which can reduce the maximum soliton velocity as measured in a real or numerical experiment. The presence of a harmonic trap causes a further reduction in the maximum observed soliton velocity. If a soliton is accelerated to its maximum velocity by an external potential, it abruptly disappears and its energy is converted into sound. Solitons may also be accelerated by inelastic collisions with other solitons, and the energy lost by the solitons is emitted as sound. This can destroy the solitons if the energy lost as sound reduces the soliton energy below the truncation point in the energy dispersion. We find that soliton collisions become increasingly inelastic as we move from the BEC to BCS regime. 

On the one hand, our results impose some limitations on experiments aimed to observe soliton oscillations and collisions in Fermi gases. Solitons on the BCS side of the resonance must be prepared close to the trap centre and accelerated gently in order to avoid their decay into sound. On the other hand, it is encouraging that solitons in fermionic superfluids behave differently to those in bosonic superfluids, and that this physics is not captured by bosonic models. The solitons in fermionic superfluids are sensitive to the fermionic degrees of freedom and so may be used as a tool to further characterize these gases.


\begin{acknowledgments}
This work has been supported by ERC through the QGBE grant. OF and JB were supported by the Marsden Fund of New Zealand (contract No. MAU0910).
\end{acknowledgments}

\bibliography{biblio}

\begin{thebibliography}{38}
\bibitem[{not()}]{note1}
\bibinfo{note}{We have checked that our soliton solutions of Eq. (1) obey the
  analytic results published in Ref.~\cite{metrento}, and hence we have
  confidence that the structure in the energy dispersions is not a numerical
  artifact.}
\end{thebibliography}

\end{document}